# A Vibrotactile Belt for Interpersonal Synchronization of Breath

Xilai Tan[1], Yan Zhang[1], Bin Zhao[1], Xiaolu Nan[1], Yuru Zhang[1,3] and Dangxiao Wang[1,2,3]

[1]*State Key Laboratory of Virtual Reality Technology and Systems, Beihang University, China*
[2]*Beijing Advanced Innovation Center for Biomedical Engineering, Beihang University, China*
[3]*School of Mechanical Engineering & Automation, Beihang University, China*

(Email: hapticwang@buaa.edu.cn)

**Abstract ---** This paper introduces a vibrotactile belt for interpersonal synchronization of breath. It can synchronize the breathing tempo of two people by transferring breathing rhythm of one user to vibration signals of another belt, where the depth of breathing is represented by the intensity of vibration. This provides a novel way of emotional connect between people. Meanwhile, this breath-sharing device may also be combined with smart devices in the future to form a one-to-many, many-to-many internet of breath, which has promising application prospects in healthcare, sports breathing guidance and other scenarios.

**Keywords: breath synchronization, emotional connect, haptics, vibration**

## 1 INTRODUCTION

Breath is one of the most important physiological activities of human beings, it can affect emotion and cognition[1]. Therefore, sharing each other's respiratory state provides a brand-new way for interpersonal emotional communication.

Most existing two-person synchronized breathing guidance devices mainly focus on visual or auditory information with no haptic feedback. However, the movement of the respiratory muscles is interrelated with haptic perception, produces lower cognitive load through passive touch[2]. Also, haptic feedback eliminates the signal translation of audio-visual input[3]. Moreover, VR or other display devices have requirements for space, not suitable for ambient respiratory sharing.

This paper proposes a wearable breath-sharing vibrotactile belt that uses vibration signals to transmit breathing rhythms bi-directionally between two individuals. Our early research mainly explores 1 question: Can two people share their breath through solely vibration signals?

## 2 SYSTEM DESCRIPTION

### 2.1 Overall Structure

The overall structure of the breath synchronization is outlined in Fig.1, and prototype of vibrotactile belt is shown in Fig.2. The control circuit, battery and 4 Linear Resonance Actuator (LRA) are integrated in one belt, to make the device wearable and lightweight. The breath signal of one user is recorded by dual-IMU measure system. We use peak-searching algorithm to detect the starting point of inspiration and expiration phase, and generate the amplitude signal accordingly. The belt receives amplitude order via Bluetooth, and produce vibrotactile stimulus to another user.

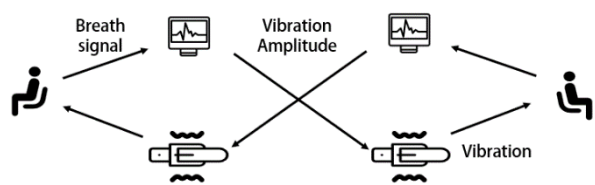

Fig.1 System diagram of breath synchronization

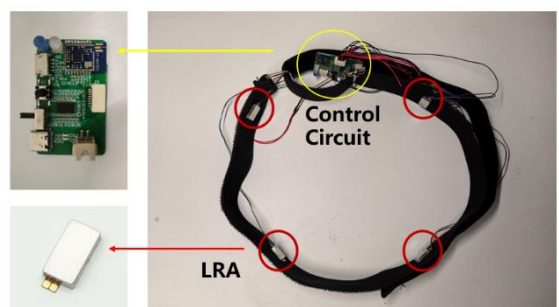

Fig.2 Prototype of vibrotactile belt

## 2.2 Dual-IMU System

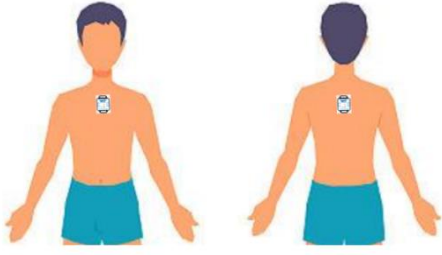

Fig.3 The attachment location of IMU

Inertial Measurement Unit (IMU) is commonly used in human motion detecting. In this study, the IMU is attached to the sternum[4] to measure respiratory motion by tracking the angular changes of the thoracic cavity to avoid soft tissue. Moreover, to eliminate the influence of body movements, another IMU is placed at the same height on the back of the body as reference:

$$a_z = a_{z,f} - a_{z,b} \quad (1)$$

Here, $a_{z,f}/a_{z,b}$ is the z-axis acceleration of the front/back IMU. By differencing the data from the measurement IMU (front) and the reference IMU (back) to calculate relative change, the interference caused by body movements can be effectively eliminated.

## 2.3 Hardware

The circuit diagram of vibrotactile belt is shown in Fig.4. The belt receives order via XY-MBA32A Bluetooth module. Linear Resonance Actuator ESA1016 is used to generate vibrotactile stimulus. The frequency is 200Hz, in the most sensitive frequency range for human skin[5].

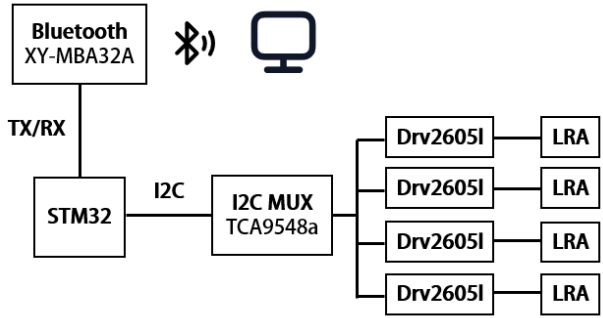

Fig.4 Circuit diagram of vibrotactile belt

## 2.4 Vibration Pattern

Based on the vibrotactile belt, we developed 3 patterns of vibration, which is shown in Fig.5. For coupled style, the intensity of vibration is synchronized with the depth of breath, while the second style is in an inversed way. The discrete style contains a sharp drop at the starting point of expiratory phase, clearly tells users when to exhale.

According to Weber's Law, the just noticeable stimulus difference is a fixed proportion of its magnitude. For vibrotactile stimulus, the Weber's fraction is about 20%. Therefore, the amplitude signal is designed to be exponential:

$$A(t) = \begin{cases} \dfrac{(14^{\frac{t}{T}} - 1)}{13} \times 100\%, & 0 < t \leq T \\ 100\%, & t > T \end{cases} \quad (2)$$

Here, $A(t)$ represents the relative intensity level. During inhale phase in coupled style, the intensity increases slower at the start to provide a smooth vibration stimulus.

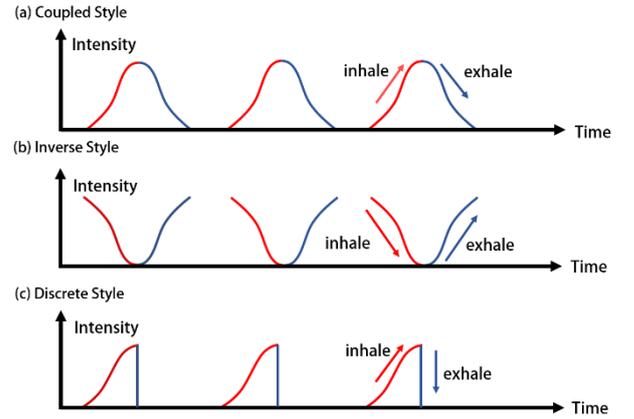

Fig.5 (a) Coupled pattern: the intensity increases at inspiratory phase, drops at expiratory phase; (b) Inversed pattern: the intensity drops at inspiratory phase, increases at expiratory phase; (c) Discrete pattern: Vibration stops at expiratory phase.

## 3 EARLY OBSERVATIONS

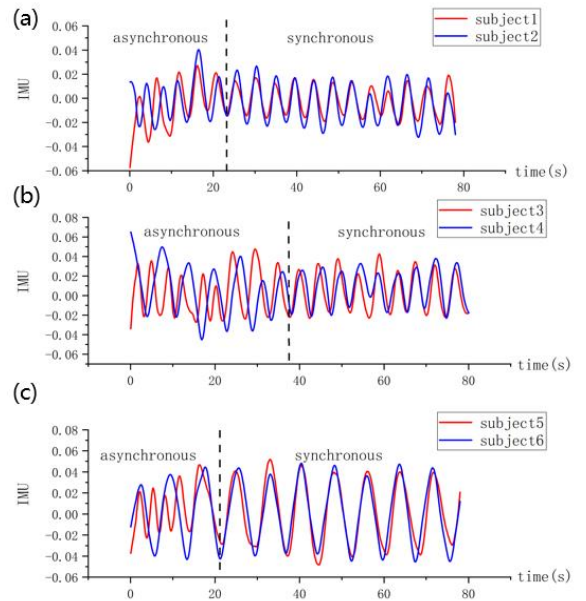

Fig.6 (a): subjects with similar breath cycle. (b), (c): subjects with different breath cycle.

Thus far, we have made early observations about the effectiveness of vibrotactile belt: Can two people synchronize their breath rhythm by vibration signals?

The result of 90-seconds experiments is shown in Fig.6. Two people sit in different room with no communication except the vibration signal of coupled style. All 3 parts of subjects synchronized their breath easily. In synchronized section, the Pearson Correlation Coefficient of IMU data is 0.85, 0.76, 0.88, which indicates a high similarity.

## 4 Prospects: Internet of Breath

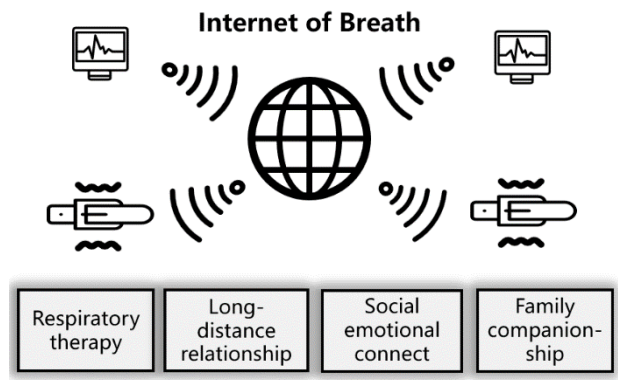

Fig.7 Outline of Internet of Breath

As shown in Fig.7, breath sharing provides a new dimension of interpersonal connection. In the future, with the help of vibrotactile breathing belt, people can upload their breath data to Internet of Breath, and connect with anyone though breath synchronizing. Vibrotactile belt will provide a pleasant and unique experience of emotional companionship in families, teams, long-distance relationships and other social emotional connections. Moreover, breath training is commonly used in rehabilitation of the elderly and postoperative patients[6]. This vibrotactile breathing belt provide a brand-new way for therapists to teach patients how to perform the exercise better.

## 5 Conclusion

This paper proposes an interpersonal breath-sharing belt based on vibration feedback that transmits breath rhythms bi-directionally between two individuals. This set of newly developed tactile modes can overcome the limitation that breathing is private and not easy to transmit, greatly reducing the distance of interpersonal communication and enhancing emotional interaction, which has a promising application in healthcare, sports breathing guidance and other scenarios.